%% file: Sharma_Pramanik_Chen_MIshra_BibIncluded.tex
\documentclass{jfm}
\usepackage{graphicx}
\usepackage{epstopdf, epsfig}
\usepackage{amsmath}
\usepackage{color}
\usepackage{tikz}
\usetikzlibrary{shapes}
\usepackage[normalem]{ulem}
\usepackage{hyperref}

\shorttitle{Reaction-induced radial fingering instability}
\shortauthor{V. Sharma, S. Pramanik, C.-Y. Chen and M. Mishra }

\title{A numerical study on reaction-induced radial fingering instability}

\author{Vandita Sharma\aff{1},
Satyajit Pramanik\aff{2}, Ching-Yao Chen\aff{3} \and 
Manoranjan Mishra\aff{1,4}
\corresp{\email{manoranjan@iitrpr.ac.in}}}

\affiliation{
\aff{1}Department of Mathematics, Indian Institute of Technology Ropar, 140001 Rupnagar, India
\aff{2}NORDITA, Royal Institute of Technology \& Stockholm University, Stockholm, Sweden
\aff{3}Department of Mechanical Engineering, National Chiao Tung University, Hsinchu, Taiwan, 30010 Republic of China
\aff{4}Department of Chemical Engineering, Indian Institute of Technology Ropar, 140001 Rupnagar, India
}

\input{macros}

\begin{document}

\maketitle

\begin{abstract}
The dynamics of $A + B \rightarrow C$ fronts  is analyzed numerically in a radial geometry. We are interested to understand miscible fingering instabilities when the simple chemical reaction changes the viscosity of the fluid locally and a non-monotonic viscosity profile with a global maximum or minimum is formed. We consider viscosity-matched reactants $A$ and $B$ generating a product $C$ having different viscosity than the reactants. Depending on the effect of $C$ on the viscosity relative to the reactants, different viscous fingering (VF) patterns are captured which are in good qualitative agreement with the existing radial experiments. We have found that for a given chemical reaction rate, an unfavourable viscosity contrast is not always sufficient to trigger the instability.  For every fixed $\Pen$, these effects of chemical reaction on VF are summarized in the $Da-\Rc$ parameter space that exhibits a stable region separating two unstable regions corresponding to the cases of more and less viscous product. Fixing $\Pen$, we determine $Da$-dependent critical log-mobility ratios $\Rp$ and $\Rm$ such that no VF is observable whenever $\Rm \leq \Rc \leq \Rp$. The effect of geometry is observable on the onset of instability, where we obtain significant differences from existing results in the rectilinear geometry. 
\end{abstract}

\section{Introduction \label{sec:Intro}}
The flows  in porous media  have been of profound interest to many researchers due to various economic and environmental processes viz., enhanced oil recovery, contaminant transport in aquifers, CO$_2$ sequestration, chromatography separation, to name a few. These displacement processes associated with a variation in the properties, viz., density, viscosity, temperature, of the flowing fluids, result in various hydrodynamic instabilities. The finger-like deformation of the interface due to the viscosity variation along the flow direction is an instability termed as viscous fingering (VF) \citep{Tan1987}. Although VF is extensively understood in non-reactive fluids, their reactive counterpart has not been well understood. Recently, the effect of chemical reaction on interfacial instabilities has attracted the focus of many researchers as the chemical reactions  interplay with these hydrodynamic flows by altering the density, viscosity of the fluid and the permeability of the porous medium \citep{DeWit2016}.

VF is studied in geometries in which one fluid displaces the other  either radially \citep{Nagatsu2007}  or rectilinearly  \citep{Maes2010}. We term them respectively as the radial and the rectilinear  geometry. Many experimental \citep{Podgorski2007,Nagatsu2007, Riolfo2012}, theoretical \citep{Hejazi2010b, Brau2017} and numerical \citep{DeWit1999, Gerard2009, Hejazi2010a, Nagatsu2011} studies have been performed where the chemical reaction suppresses, enhances or triggers miscible VF by varying the spatiotemporal evolution of the concentration of the fluids. \citet{DeWit1999} studied the interactions of reaction and VF in a porous medium through numerical simulations and observed a new droplet formation mechanism. Experimental investigation of the effects of an infinitely fast reaction on VF in a radial Hele-Shaw cell reveal that suitable choices of reactants and a chemical reaction can be an efficient control parameter for miscible VF \citep{Nagatsu2007}. These experimental results are supported with numerical simulations in the rectilinear geometry \citep{Nagatsu2011}. For a finite $Da$, chemical reaction-induced instabilities are captured from radial experiments when a more viscous reactant displaces a less mobile reactant, and are supported qualitatively using nonlinear simulations in rectilinear model \citep{Riolfo2012}. In parallel, \citet{Podgorski2007} experimentally studied instability borne out solely by reaction in a radial Hele-Shaw cell, by considering reactants having same viscosity and observed a variety of fingering instabilities. Subsequently, numerical simulations are performed in rectilinear geometry for viscosity-matched reactants, when brought in contact, generating a more viscous product and thus resulting a viscosity maximum along the flow direction \citep{Gerard2009}, wherein by incorporating different diffusivity of the reactant fluids into their numerical model, authors succeed to explain the asymmetry in the fingers observed from the experiments of \citet{Podgorski2007}. Recently, \citet{Brau2017} presented a combined study of both theory and experiments in a radial geometry without fingering instability. To the best of our knowledge, the numerical simulations of reactive VF in radial geometry are not available in literature. This can be attributed to the singularity in the velocity field in the radial flow, which in the absence of interfacial instabilities, is $\propto 1/r$, where $r$ is the radial distance from the source. On the other hand, a rectilinear displacement in a rectangular domain is free from the singularities in the velocity field. Thereby making rectilinear geometry a preference for non linear simulations, whereas experiments are predominantly done in a radial Hele-Shaw cell.

The radial advection of reacting species has received growing attention due to it's wide spectrum of application ranging from chemical garden \citep{Haudin2014} to the growth of bacterial colonies \citep{Lega2007}, to the study of an active fluid interface \citep{Nagilla2018}, to infectious disease spreading \citep{Brockmann2013}. Motivated by the broad applications of radial transport in reactive systems, we perform non-linear simulations of reaction induced radial VF. This forms the novelty of our work along with paving a way for good quantitative comparison between experiments and numerical works. In this paper we answer two important questions: Is it the chemical reaction or the viscosity contrast between the fluids that dominates reactive VF? What are the geometric effects on this fingering dynamics?  For this, we consider the reactants to be of equal viscosity as in \citet{Podgorski2007}, and investigate the effects of chemical reaction on fingering instability in terms of two non-dimensional parameters (a) Damk\"ohler number ($Da$) and (b) log-mobility ratio ($\Rc$). We find that indeed the geometry plays a crucial role on the dynamics. Further, we report the existence of two critical mobility ratios for a given $Da$ for the occurrence of VF. Consequently, there exists a stable region in $Da-\Rc$ parameter space separating two unstable regions. Our results improve the understanding of the effect of chemical reactions on miscible VF and this paves a way to use chemical reactions to control miscible VF. \section{Mathematical formulation and numerical method}\label{sec:model}
\bfig
\centering
\includegraphics[scale=0.4]{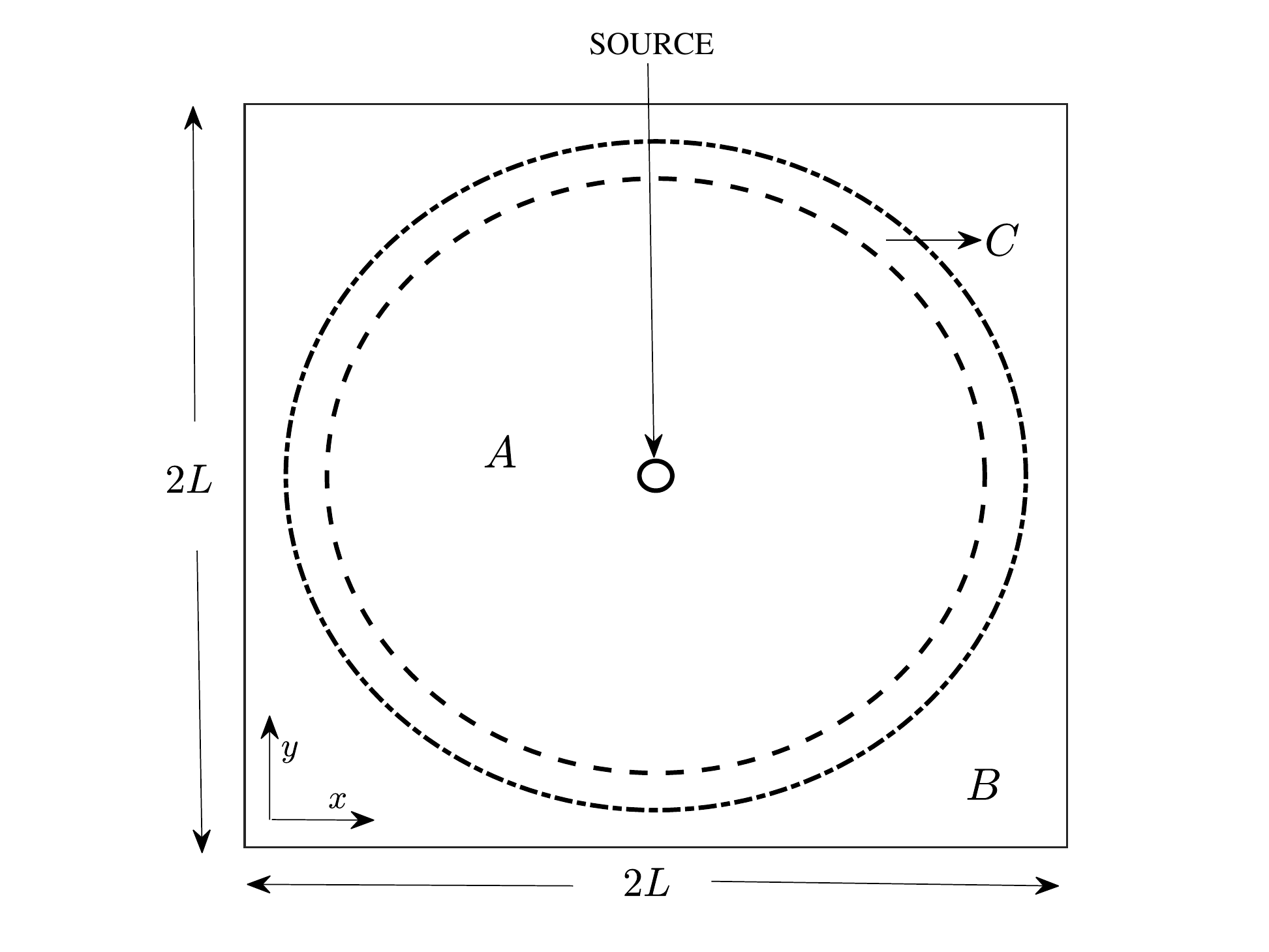}
\caption{Schematic showing   $A + B \xrightarrow{k} C$ in a radial source flow. Dashed curve represents miscible interface between \emph{fluid A} and \emph{C}, while dash-dotted curve represents the same between \emph{fluid B} and \emph{C}. The region between these curves shows the ring of product.}
\label{fig:Schematic} 
\efig
We consider a single phase fluid flow in a two-dimensional homogeneous, isotropic, saturated porous medium. We further assume that there exist two solutes $A$ and $B$ dissolved in a solvent and they produce a new solute $C$ by a second order chemical reaction 
$
A + B \xrightarrow{k} C.
$
Without a loss of generality we call the solution of the solute $A$ in the solvent as \emph{fluid A}, and similarly, \emph{fluid B} and \emph{fluid C} are defined as the solution of the solute $B$ and $C$ respectively in the solvent. As shown schematically in figure \ref{fig:Schematic}, the \emph{fluid A} is injected radially from a point source at the center of the porous medium at a constant flow rate per unit depth, $Q$, to displace the \emph{fluid B} that initially saturates the porous medium.  We assume that the three fluids are incompressible, Newtonian, and neutrally buoyant--the gravitational forces are neglected in our model.  The governing equations include the Darcy's law \citep{Chen2008, Gerard2009} for the fluid flow in a homogeneous porous medium:
\bseq
\beqn
\label{eq:Continuity_dim}
& & \tdiv{\u} = 0, \\ 
\label{eq:Darcy_dim}
& & \u = - \frac{\k}{\m\bra{\c}} \tgrad{\p}, 
\eeqn
\eseq
wherein $\u = (\wt{u}, \wt{v})$ and $\p$ respectively denote the Darcy's velocity and the fluid pressure, $\k$ is the constant permeability of the porous medium. $\a, \; \b, \; \c$, denote the concentration of the two reactants ($A, B$) and the product ($C$), respectively, and their mass balance satisfy the following system of coupled non-linear partial differential equations (PDEs) \citep{Gerard2009, DeWit2016}: 
\bseq
\beqn
\label{eq:CDR_A_dim}
& & \frac{\pa \a}{\pa \tt} + \bra{\u \bcdot \wt{\bnabla}} \a = \tdiv{\bra{\DA \tgrad{\a}}} - k \a\b, \\
\label{eq:CDR_B_dim}
& & \frac{\pa \b}{\pa \tt} + \bra{\u \bcdot \wt{\bnabla}} \b = \tdiv{\bra{\DB \tgrad{\b}}} - k \a\b, \\
\label{eq:CDR_C_dim}
& & \frac{\pa \c}{\pa \tt} + \bra{\u \bcdot \wt{\bnabla}} \c = \tdiv{\bra{\DC \tgrad{\c}}} + k \a\b, 
\eeqn
\eseq
where $\DA, \; \DB, \; \DC$ are the diffusion coefficients of $A, \; B, \; C$ in the solvent, and $k$ is the reaction rate constant.
The dynamic viscosity of the reactant fluids, \emph{fluid A} and \emph{fluid B}, are the same, and they differ from that of the product, \emph{fluid C}. The viscosity of fluids are modeled as an Arrhenius relation similar to the earlier works on reactive VF \citep{Hejazi2010b}. 

\subsection{Non-dimensionalisation}
To be able to explicitly explore the effect of chemical reaction rate and the diffusion of the species ($A, \; B, \; C$) on the dynamics, we render for non-dimensionalization the characteristic length, time, velocity, solute concentration, viscosity, and pressure, as 
\bseq
\beqn
\label{eq:scaling1}
& & \bx = \frac{\wt{\bx}}{\sqrt{Q t_f}}, \qquad t = \frac{\tt}{t_f}, \qquad \bu = \frac{\u}{\sqrt{Q/t_f}}, \\ 
\label{eq:scaling2}
& & \bra{a, b, c} = \frac{\bra{\a, \b, \c}}{a_0}, \qquad \mu = \frac{\m}{\ms}, \qquad p = \frac{\p}{Q \ms/\k}. 
\eeqn
\eseq
Here, $t_f$ is the final duration of the fluid injection. The characteristic viscosity $\ms$ is defined below. 
The resultant initial value problem written in dimensionless form satisfies coupled nonlinear partial differential equations: 
\bseq
\beqn
\label{eq:Darcy}
& & \divg{\bu} = 0, \quad
\bu = - \frac{1}{\mu(c)} \grad{p}, \quad 
\mu(c) = \exp{\bra{\Rc c}}, \\ 
\mbox{and} \nn \\ 
\label{eq:CD_A}
& & \frac{\pa a}{\pa t} + (\bu \bcdot \bnabla) a = \frac{1}{\Pen} \lap a - Da \; a b,\\
\label{eq:CD_B}
& & \frac{\pa b}{\pa t} + (\bu \bcdot \bnabla) b = \frac{1}{\Pen} \lap b - Da \; a b, \\
\label{eq:CD_C}
& & \frac{\pa c}{\pa t} + (\bu \bcdot \bnabla) c = \frac{1}{\Pen} \lap c + Da \; a b, 
\eeqn
\eseq
associated with the initial conditions 
\bseq
\beqn
\label{eq:IC_u}
& & \bu = \bra{ x \sqbra{2 \upi \bra{x^2 + y^2}}^{-1}, \; y \sqbra{2 \upi \bra{x^2 + y^2}}^{-1} }, \\ 
\mbox{and} \nn \\ 
\label{eq:IC_A}
& & a(x, y, t = 0) = \left\{
\begin{array}{ll}
1, & 0 < x^2 + y^2 \leq d^2/4 \\ 
0, & \mbox{Otherwise} \\
\end{array} 
\right. , \\ 
\label{eq:IC_BC}
& & b(x, y, t = 0) = 1 - a(x, y, t = 0), \qquad c(x, y, t = 0) = 0 \quad \forall x,y, 
\eeqn
\eseq
wherein $d$ is the dimensionless diameter of a circular region. 
The \emph{three} dimensionless parameters appearing in equations \eqref{eq:Darcy}--\eqref{eq:CD_C} are 
\beq
\label{eq:dimless_para}
\Rc = \ln \bra{\frac{\mu_C}{\ms}}, \qquad Da = \frac{t_f}{1/\bra{k a_0}}, \qquad \Pen = \frac{Q}{D}, 
\eeq
where $\ms = \mu(a + b = 1, c = 0)$, $\mc = \mu(a + b = 0, c = 1)$. 
Although, in general, the diffusion coefficients of the three species $A, \; B, \; C$ are different and not necessarily constant, we assume these to be identical and equal to a constant $D$. This simplification allows to focus on the understanding of the influence of a simple chemical reaction on the fingering instability in an otherwise stable miscible displacement. The P\'eclet number ($\Pen$) definition is similar to that used in experiments by \citet{Nagatsu2007}, which is a consequence of the scaling used, allowing a direct comparison with experiments. While 
$\Pen$ determines the relative importance of the advection to the diffusion of the solute, the Damk\"ohler number, $Da$, is defined as the ratio of the advective time scale to the reaction time scale. A higher $Da$ corresponds to a smaller reaction time (a faster reaction rate) and hence the product is formed fast. Thus for every fixed $\Pen$, varying the two dimensionless parameters $Da$ and $\Rc$, we explore the effect of chemical reaction on fingering dynamics. 
Equations \eqref{eq:Darcy}--\eqref{eq:CD_C} are solved using a hybridization of compact finite difference and the pseudo-spectral method for the spatial derivative, while fully explicit third order Runge-Kutta method is used for the time integration \citep[and refs. therein]{Chen2008}.  Similar numerical scheme has been quite effective with the non-reactive counterparts where good  qualitative agreement with the theoretical results are available in literature. To name a few,  \citet{Chen1998a} reported a good agreement with  the growth  rate on miscible interface, while for the immiscible  fluids, the number of fingers triggered \citep{Chen2011,Chen2014} are found consistent with the theoretical results. We briefly discuss  our method for the reactive miscible fluids  below. 

\subsection{Numerical method} \label{Num_Meth} 

Numerical computations are performed in a square box $\Omega = \sqbra{-L, L} \times \sqbra{-L, L}$. In the absence of any viscosity gradient, 
 the radial flow is a potential flow with velocity vector, $\upot = \bra{ x \sqbra{2 \upi \bra{x^2 + y^2}}^{-1}, \; y \sqbra{2 \upi \bra{x^2 + y^2}}^{-1} }$. However, this potential flow is perturbed as chemical reaction induces a viscosity variation at the miscible interface between the two reactants. In order to capture this deviation of the velocity from the potential flow velocity $\upot$, we decompose the velocity vector at any time into two parts, 
 \beq
 \label{eq:u_decompose}
 \bu = \upot + \urot.
 \eeq
 The singularity in the potential flow at the origin is regularized by introducing a Gaussian source of core $\sigma$ as in \citet{Chen2008}:  
\beq
\label{eq:upot_regular}
\upot = \bra{ \frac{ \bra{ 1 - e^{-\bra{x^2 + y^2}/\sigma^2} } x}{ 2 \upi \bra{x^2 + y^2} }, \frac{ \bra{ 1 - e^{-\bra{x^2 + y^2}/\sigma^2} } y}{2 \upi \bra{x^2 + y^2}} },
\eeq
here $\sigma \leq d/2$, the radius of the circle initially occupied by reactant $A$. We take $d=0.15$ in all the simulations.
Introducing a stream function $\psi$ such that $\urot = \bra{ \pa \psi/\pa y, -\pa \psi/\pa x }$, we write the Darcy's law, equation \eqref{eq:Darcy}, by eliminating the pressure, as 
\refstepcounter{equation}
$$
\label{eq:Poisson}
 \lap \psi = - \om, \quad
 \label{eq:Vorticity}
 \om = \Rc \sqbra{ v \bra{\pa c/\pa x} - u \bra{\pa c/\pa y} }.
  \eqno{(\theequation{\mathit{a},\mathit{b}})}\label{eq35}
$$

We discretize the advection and diffusion terms in equations \eqref{eq:CD_A}--\eqref{eq:CD_C} using \emph{fourth order} and \emph{sixth order} finite difference formulae \citep{Lele1992} to obtain semi-discretized equations
\refstepcounter{equation}
$$
\label{eq:CD_A_IVP}
\pa_t a = f(a,b,c), \qquad
\label{eq:CD_B_IVP}
\pa_t b = g(a,b,c), \qquad
\label{eq:CD_C_IVP}
\pa_t c = h(a,b,c),
\eqno{(\theequation{\mathit{a},\mathit{b},\mathit{c}})}\label{eq36}
$$
where $f, \; g, \; h$ are the discretized version of the advection-diffusion-reaction operators corresponding to their continuous form in equations \eqref{eq:CD_A}--\eqref{eq:CD_C}. No flux boundary conditions are used for the solute concentrations and $\psi=0$ across all four boundaries. The equations \eqref{eq:CD_A_IVP} are associated with the initial conditions \eqref{eq:IC_u}--\eqref{eq:IC_BC}. For numerical convergence, 
the step-like initial condition is regularized as follows. Equation \eqref{eq:IC_A} tells that the reactant $A$ is initially contained within a circle of radius $d/2$, which is equivalent to a dimensionless diffusive time $t = d^2/8$. Thus, the axisymmetric base state solution of \cite{Tan1987} at $t = d^2/8$ is used as an initial condition for $a$. We introduce a random disturbance of amplitude $\xi \sim \textit{O}\bra{10^{-2}}$ around the points where $a = 0.5$, so that concentration at those points is given as
$ a = 0.5 + \xi \sin(2 \upi N),$
where $N$ is a random number between $0$ and $1$. The same set of random numbers is used for all the simulations. The initial condition for $a$ subtracted from unity gives the initial condition for $b$. 

The system of equations \eqref{eq:CD_A_IVP} is solved using fully explicit third order Runge-Kutta method with adaptive time stepping that satisfy the Courant-Friedrichs-Lewy condition. 
Sixth-order compact finite difference approximations are used in equation \eqref{eq:Poisson} to compute $\om$. A fast and efficient solution of the Poisson equation in \eqref{eq:Poisson} is achieved using a hybridization of compact finite difference and the pseudo-spectral method. We employ a Galerkin-type discretization in the $x$-direction and a sixth order compact finite difference in the $y$-direction for $\psi$. 
The simulations are stopped as soon as the concentration near the boundary becomes more than $0.1$ for $a$ or less than $0.9$ for $b$. In this study, we take $\Omega=[-0.75,0.75] \times[-0.75,0.75]$  with $512 \times 512$ grid points.

\section{Results and discussions} 
\bfig
\centering
\includegraphics[scale=0.5]{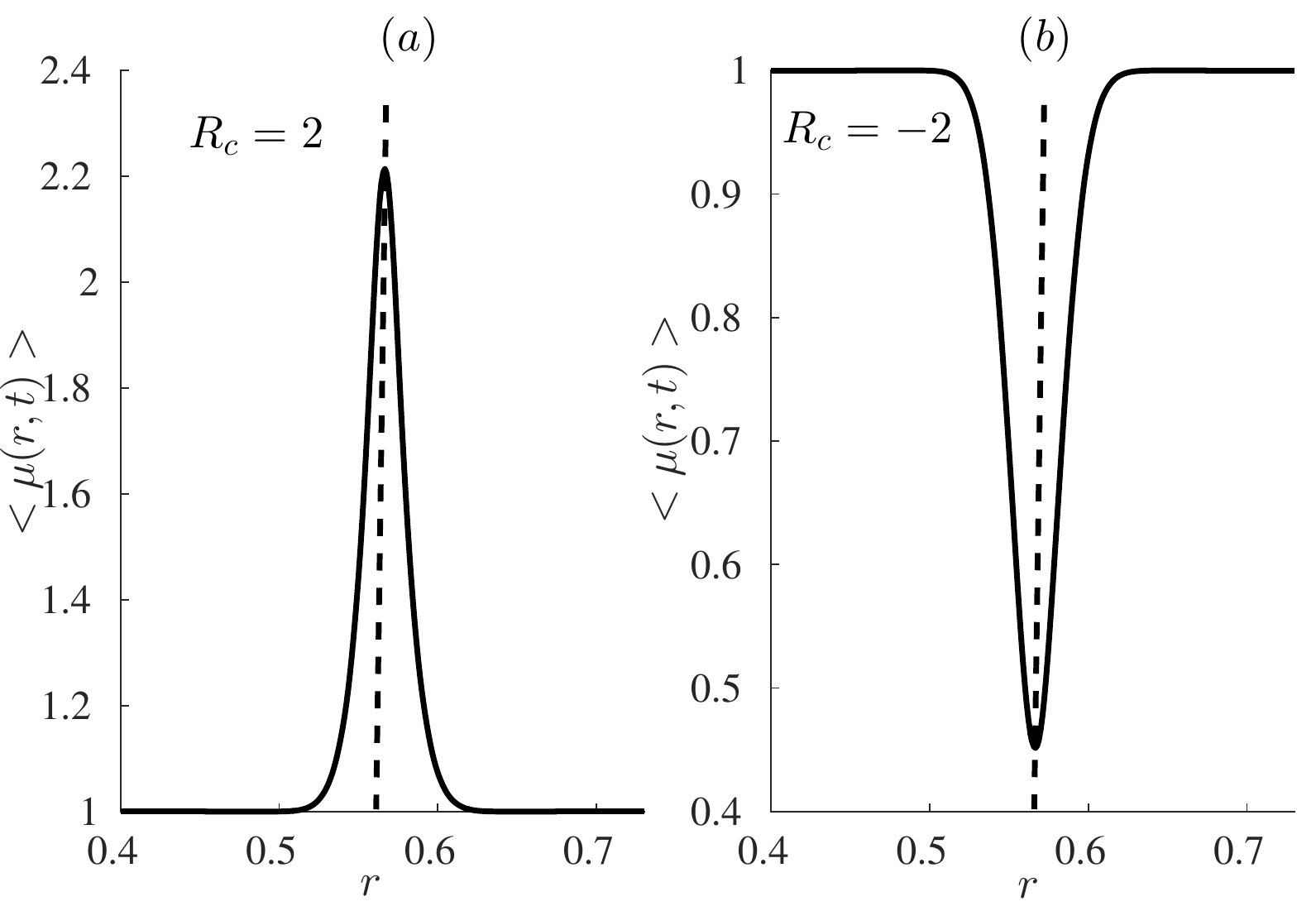} 
\caption{Viscosity profile across the interface (dashed line) along the radial direction for $Da=100, t=1$.}
\label{fig:visc} 
\efig
As mentioned in \S \ref{sec:model}, the dynamic viscosities of \emph{fluid A} and \emph{fluid B} are identical, which corresponds to a stable displacement in the absence of chemical reaction or when the product viscosity is identical to that of the reactants. How does a chemical reaction alter this stable displacement if \emph{fluid C} formed due to the chemical reaction has a viscosity different from the reactants? The product thus generated by the chemical reaction introduces a non-monotonicity in the viscosity profile as shown in figure \ref{fig:visc}. Understanding the effect of the non-monotonicity can be of great help in controlling the instability. In figure \ref{fig:DA100}, we show the temporal evolution of both the more as well as less viscous product generated by the chemical reaction. 
\bfig
\centering
\includegraphics[scale=0.85]{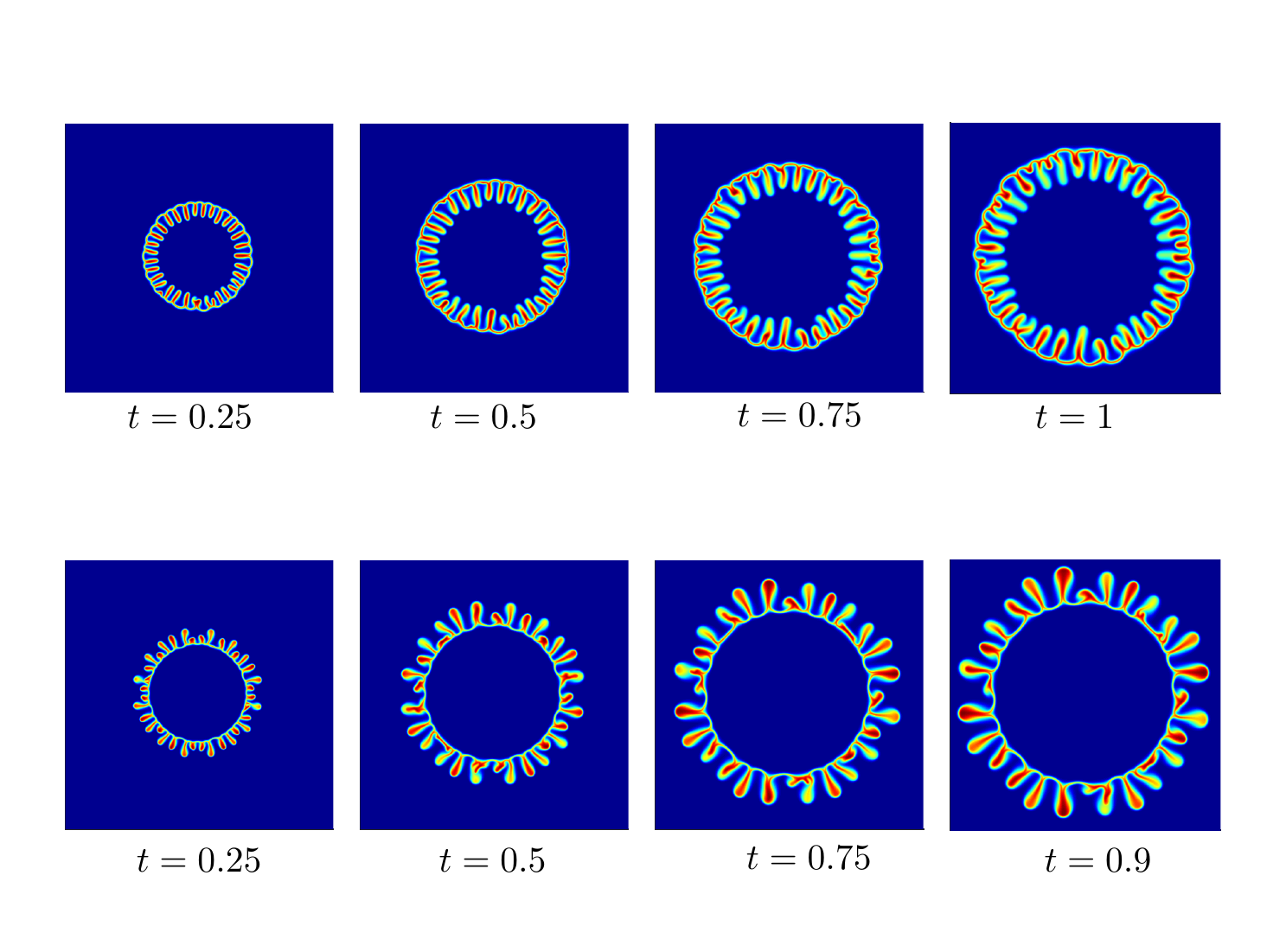}
\caption{Temporal evolution of  the concentration of product $C$ for $Da=100, Pe = 3000, \Rc=7$ (first row) and $\Rc=-7$(second row).}
\label{fig:DA100} 
\efig

$\Rc > 0$ corresponds to the case when the product viscosity is larger than the reactants. In this case, the viscosity $\mu(c)$ attains a local maximum at the reaction zone as shown in figure \ref{fig:visc}(\textit{a}). For such viscosity profiles, the miscible interface between \emph{fluid A} and \emph{fluid C} is anticipated to be hydrodynamically unstable, whereas the miscible interface between \emph{fluid C} and \emph{fluid B} resists the growth of the fingers away from the source. As a consequence, the fingers form inside the ring of the product observable in the first row of figure \ref{fig:DA100}. Further, the tip of the outward growing fingers into the interface between \emph{fluid C} and \emph{fluid B} is wider than the root of the corresponding fingers. This is in good qualitative agreement with the viscosity increasing experiments \citep{Nagatsu2007, Riolfo2012} as well as the fan state fingering instability \citep{Podgorski2007}.
\bfig
\centering
\includegraphics[scale=0.82]{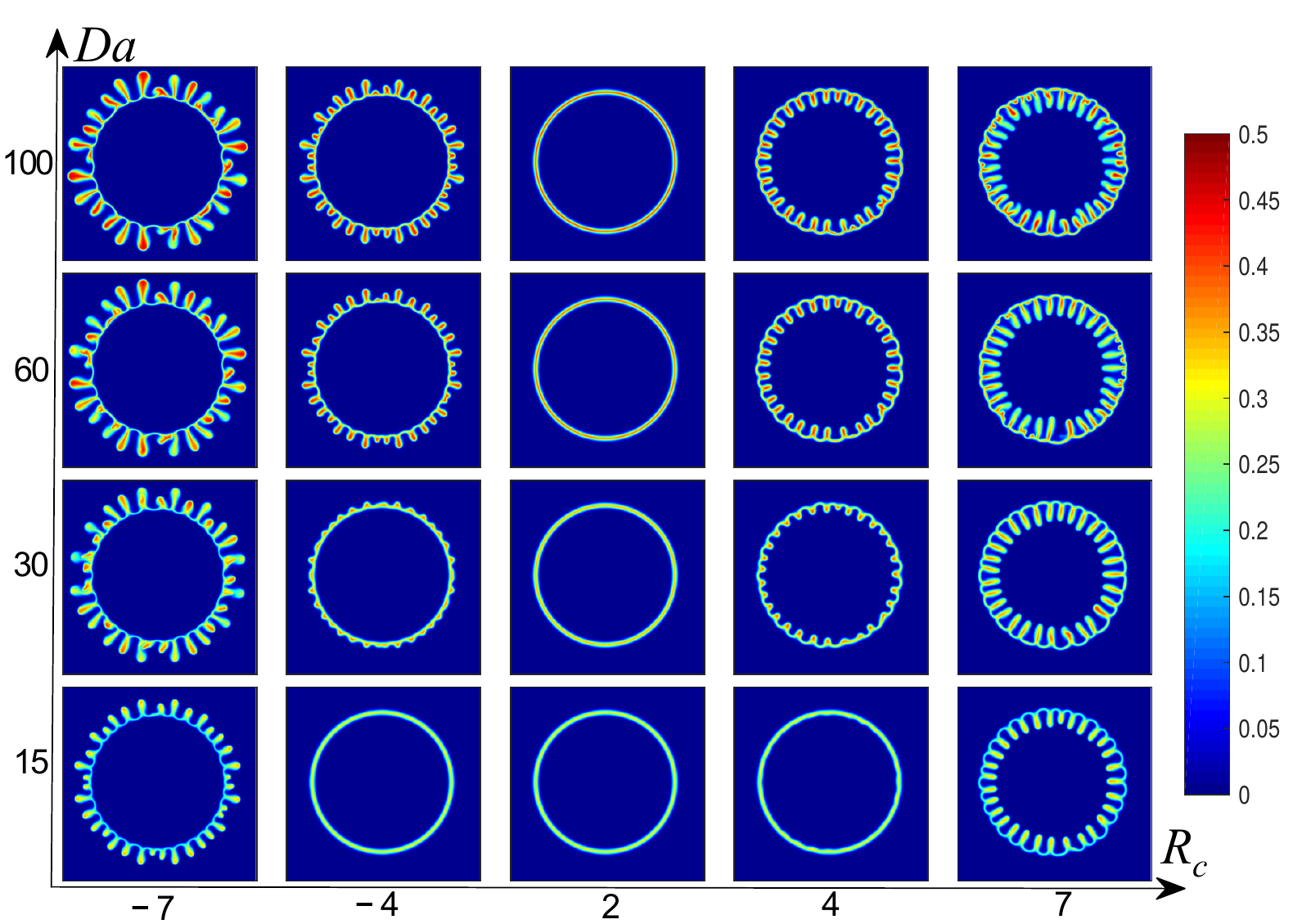} 
\caption{Temporal evolution of  the concentration of product $C$ for various $Da$ and $R_c$ and $Pe = 3000$ at $t=0.9$.}\label{fig:DA_RC} 
\efig

On the other hand, when the viscosity of the product is less than both the reactants, i.e., $\Rc < 0$, the interface between \emph{fluid A} and \emph{fluid C} resists the growth of the fingers formed at the interface between \emph{fluid C} and \emph{fluid B}. The fingers develop at the outer interface of the ring of product and grow in the direction of the bulk fluid motion. Consequently, for larger $Da$, the fingers quickly reach the boundary of the domain. In particular, for $\Rc = -7$ and $Da =100$, the outward fingers reach domain boundary before $t = 1$ and hence in second row of figure \ref{fig:DA100}, the temporal evolution is shown upto time $t = 0.9$ only. The fingering patterns for $\Rc<0$  are also in good qualitative agreement with existing experiments in the radial geometry \citep{Riolfo2012}. A comparison of the density plots for positive and negative $\Rc$, reveals that for $\Rc<0$, the fingers are narrower away from source and hence cover smaller area, another feature, in good agreement with the experiments in which reaction decreases the viscosity of the fluids locally \citep{Nagatsu2007}.  

The qualitative features depend on the sign of $\Rc$ but the intensity and the interaction of fingers depend upon the reaction rate too. Our numerical simulations reveal a variety of fingering patterns, which emerge due to a coupled effect of chemical reaction and hydrodynamics. In figure \ref{fig:DA_RC}, we plot the reaction-induced fingering dynamics at $t=0.9$ for $\Pen = 3000$ and a wide range of $Da$ and $\Rc$, including both the cases of more and less viscous product forming reactions. For $ \Pen=3000, Da=60,100, \Rc=-7$, the fingers reach the boundary  after $t=0.9$, hence all the plots are shown at $t=0.9$. For every $Da$, an increasing finger interaction and vigorous fingering instability is evident as $\modu{\Rc}$ increases. This is in qualitative agreement with the linear stability results of \citet{Hejazi2010b} for viscosity matched reactants. Depending upon $Da$, a stable displacement even for an unfavourable viscosity contrast between the displacing and displaced fluid is a novel finding of our work. Also for $Da=15, \Rc=4$ bumps are visible which are absent for $\Rc=-4$. Thus, reaction producing more viscous product is more unstable. We explain this in the next subsection \S \ref{sec:quant}. 

 \subsection{Quantitative analysis} \label{sec:quant}
For the quantitative analysis, we make the transformations from $(x, y)$ coordinate to $(r, \theta)$ coordinate.
The conversions are made from $r= \sigma$ to $r=L$ in order to avoid the effect of Gaussian source, if any, near the origin. So the domain in the $(r,\theta)$ coordinate is $[\sigma, L] \times [ 0,2 \upi)$. We carry the quantitative analysis in the $(r,\theta)$ plane in the next two subsections \S \ref{sec:Onset} and \S \ref{Finger_length}.
\subsubsection{Onset of instability}
\label{sec:Onset}
Interfacial length is the length of the product ring for a stable displacement and the length of the curve along the fingering pattern for an unstable displacement. In order to quantify the interfacial length, we define it in the $(r, \theta)$ plane as,
\beq
\label{eq:Interfacial_length}
 I(t) = \int_{\sigma}^{L} \int_{0}^{2\upi} \sqrt{\bra{ \frac{\pa c}{\pa r} }^2 + \frac{1}{r^2} \bra{ \frac{\pa c}{\pa \theta} }^2 } \; r \mathrm{d}\theta \mathrm{d}r.
 \eeq 
 \bfig
\centering
\includegraphics[scale=0.75]{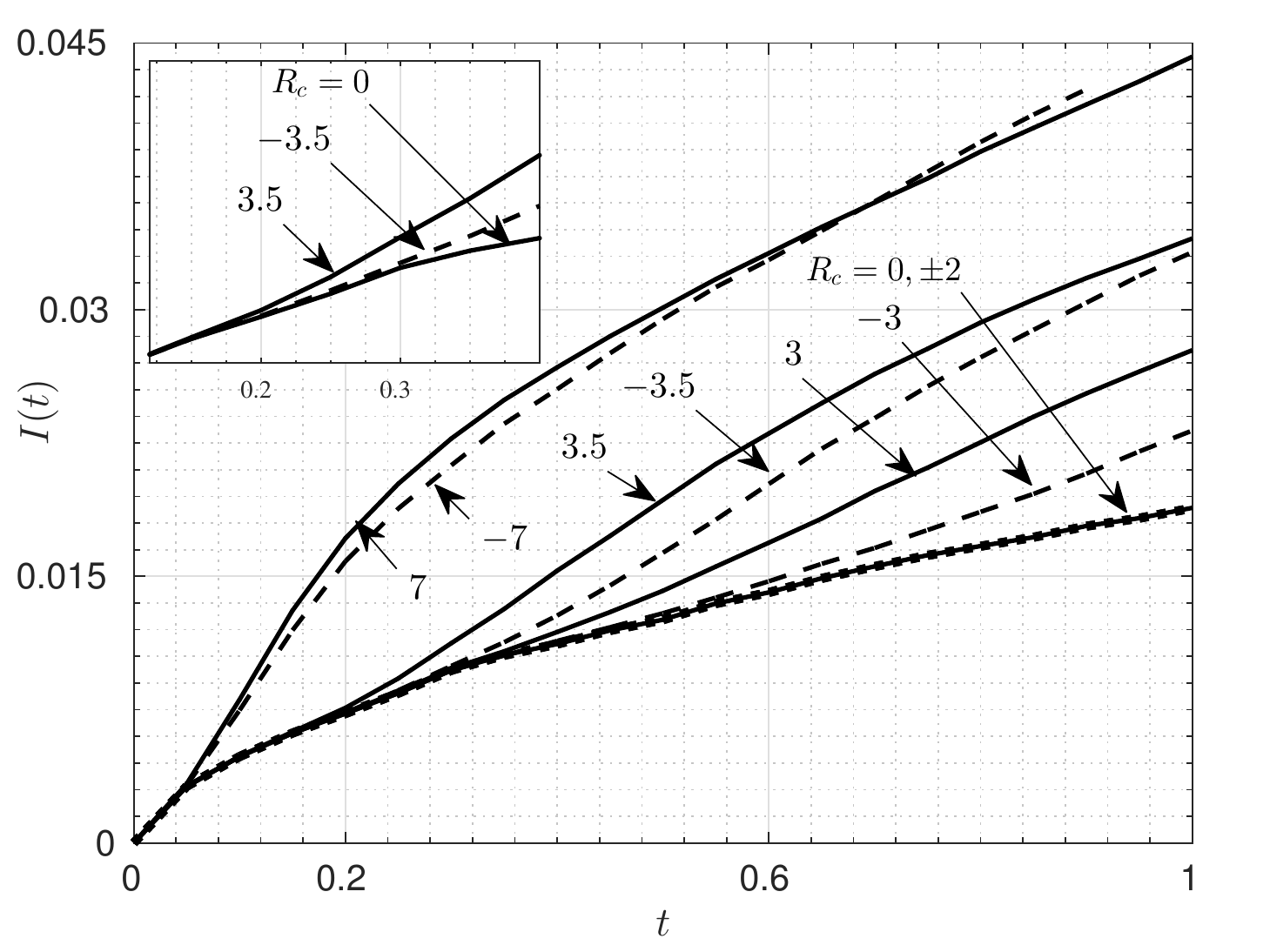} 
\caption{ Interfacial length as a function of time for $ Da = 100, \Pen = 3000 $ and various $\Rc$. Inset shows zoomed plot around time $t \in (0.04,0.44)$ in order to highlight the onset time for $\Rc=3.5$ and $ \Rc=-3.5$. Clearly, the instability is triggered late for $\Rc=-3.5$. }
\label{fig:interfacial} 
\efig
The numerical integration is carried out using trapezoidal rule. In the $(r,\theta)$ plane, $dr$ is fixed by the spatial discretisation in $(x,y)$ plane as $ dr=\sqrt{dx^2+dy^2}$ but $d\theta = tan^{-1}(dy/dx)$ results in a poor discretisation. So, we fix $d\theta=10^{-4}$  and carry out the conversions from $(x,y)$ to $(r,\theta)$ plane. It is verified that decreasing $d\theta$ further does not qualitatively and quantitatively change our results. $I(t)$ normalised with $d\theta$ is shown in figure \ref{fig:interfacial}.

I(t) can be used as a tool to gain an insight into the onset of instability. The instant $I(t)$ for $\Rc \neq 0$ departs from that for $\Rc=0$, is the time when instability triggers and is termed as the onset time. It is observable from figure \ref{fig:interfacial} that difference between onset time for $\Rc<0$ and corresponding positive $\Rc$ decreases as we increase $\modu{\Rc}$ and ultimately the onset time equals. This is because, before fingering occurs, the product is formed in a ring and the difference in onset time is attributed to the fluid velocity at the unstable interface of the ring of product. For $\Rc < 0$, fingers are formed at the outer interface of the ring, where the fluid velocity of the unperturbed flow is smaller than the velocity at the inner interface of the product ring. Therefore the advective force, which is responsible for the fingering instability, is weaker in the former case. As $\modu{\Rc}$ increases, the onset of instability decreases and it is independent of width of the ring. Hence for a given $Da$,  onset time for $\Rc<0$ is always less than or equal to, that for corresponding positive $\Rc$. This is a contradiction to the linear stability analysis in a rectilinear geometry  \citep{Hejazi2010b} and can be attributed to the spatially varying velocity of the potential flow. Thus the geometry used for the non linear simulations do play a role in the dynamics and can lead to different results. Also, $I(t)$ for many non-zero $\Rc$ coincides with that for $\Rc=0$ for all time. This restates the existence of stable flows for a range of $\Rc$ despite an unfavorable viscosity contrast. 
\subsubsection{Effect of barrier}
 \label{Finger_length}
It is evident from figure \ref{fig:visc} that the non monotonicity introduced in the viscosity profile by the chemical reaction results in a hydrodynamic stable interface where a more viscous fluid displaces a less viscous one. The stable interface hinders the fingers growing at the other interface. This is observable from the density plots in figure \ref{fig:DA_RC} where the fingers either only grow inwards for $\Rc >0$ or outwards for $\Rc<0$. In order to quantify the effect of the barrier, we define the angular average concentration as 
\beq
\abra{c(r,t)} = \bra{2\upi}^{-1} \int_0^{2\upi} c(r, \theta, t) \mathrm{d} \theta.
\eeq
\bfig
\centering
\includegraphics[scale=0.65]{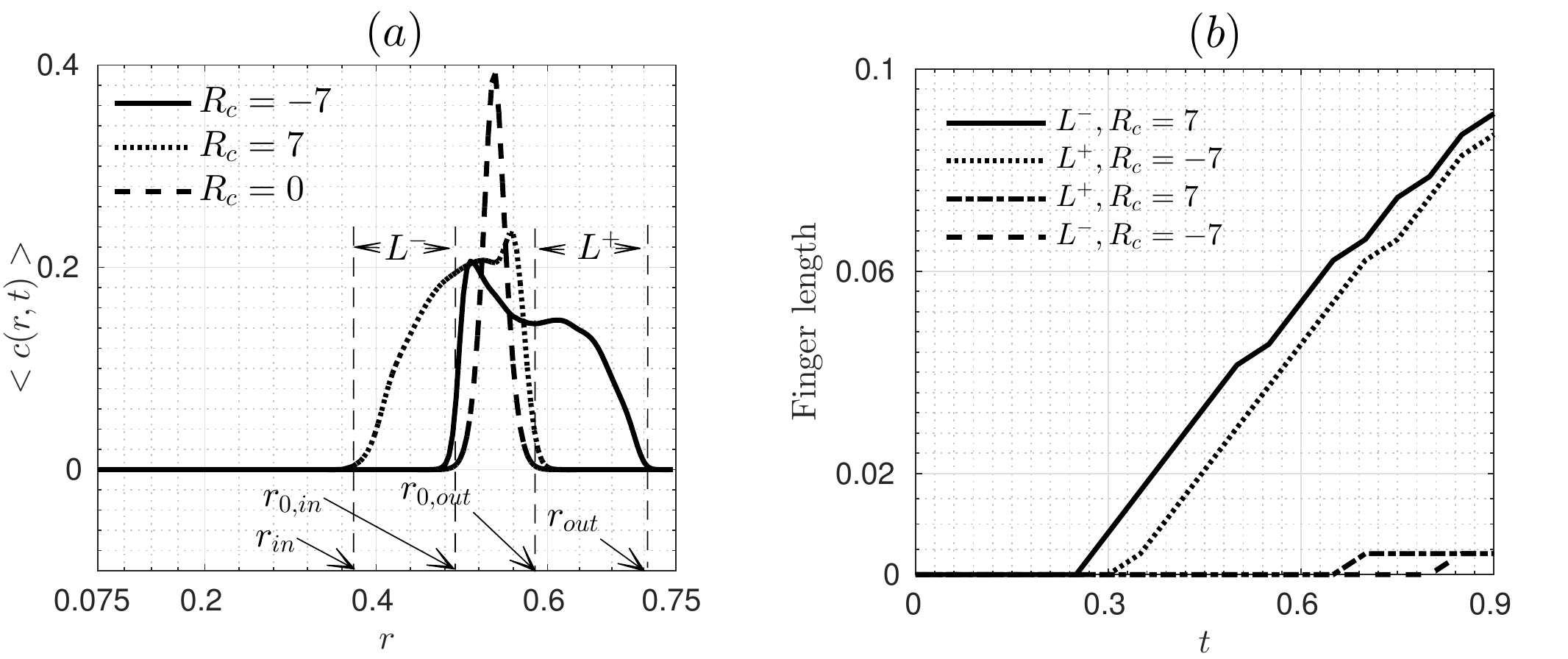} 
\caption{(\textit{a}) Angular average concentration, $\abra{c(r,t)}$ at $t = 0.9$ for $Da=100, \Pen=3000$ and various $\Rc$.  $r_{0,in}$, $r_{0,out}, r_{in}$ and $r_{out} $ used to compute $ \Lp$ and $\Lm$ are also shown. (\textit{b}) Outward and inward finger length respectively for $\Rc =-7 $ and $7$ for $Da = 20$.}
\label{fig:finger} 
\efig
  \bfig
\centering
\includegraphics[scale=0.65]{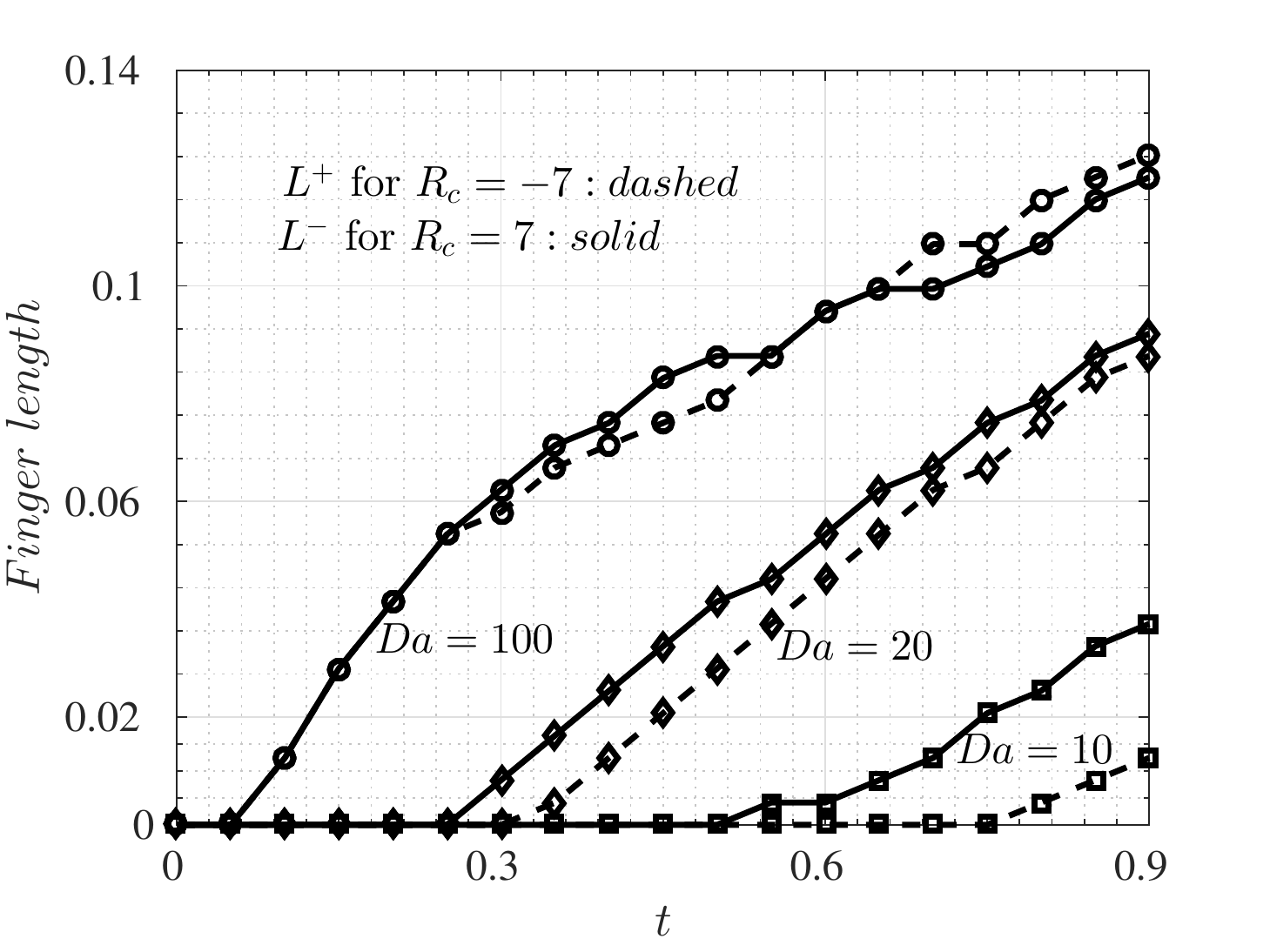} 
\caption{Outward and inward finger length respectively for $\Rc =-7 $ and $7$ for $Da = 100$, $20$ and $10$.}
\label{fig:finger_DA} 
\efig
This quantification is possible only due to the transformation from $(x, y)$ to $(r, \theta)$ plane. Figure \ref{fig:finger}\textit(a) shows $<c(r,t)>$ at $t=0.9$ for various $\Rc$ and $Da=100, \Pen=3000$. The bumps towards the left for $\Rc=7$ and towards right for $\Rc=-7$ show that fingers predominantly grow inwards and outwards, respectively for positive and negative $\Rc$. Further, we use $<c(r,t)>$ to find the length of fingers.
We define the fingers growing at the outer interface of product ring as outward fingers while the fingers at inner interface are defined as inward fingers. To analyse the effect of barrier induced by chemical reaction, we define the outward and the inward finger length as
 \beq
 \label{eq:finger_length}
 \Lp = |\roo-\ro| \text{ and }  \Lm= |\r-\rin|, 
  \eeq
respectively. Here, $r_{0,in}$ and $r_{0,out}$ are the radii when $\abra{c(r,t)} \approx 10^{-4}$ for $\Rc=0$  as shown in figure \ref{fig:finger}(\textit{a}) and $r_{in}$ $\&$ $\ro$ are the corresponding values for $\Rc \neq 0$. Figure \ref{fig:finger}(\textit{b}) shows $\Lp$ is larger (smaller) than $\Lm$ for negative (positive) $\Rc$, thus capturing the effect of the barrier introduced by the non-monotonicity, on the finger length.

We capture the effect of chemical reaction on the finger length in figure \ref{fig:finger_DA}. It is clear that reaction producing more viscous product is more unstable as length of fingers for positive $\Rc$ is always more than that for corresponding $\Rc<0$, fixed $Da$. This is due to the radially varying potential component of velocity which delays the onset and consequently reduces finger length for $\Rc<0$. Further, as $Da$ increases, the two finger lengths are found to approach each other, with $\Lm$ ultimately surpassing $\Lp$. This is attributed to the barrier in the direction of bulk fluid motion for  positive $\Rc$ which hinders the growth of fingers.
\subsection{Critical $R_c$ for the fingering instability}\label{subsec:Rc_crit} 
Above discussions reveal that for each $Da$, a minimum viscosity contrast is required for the fingering instability to occur. To determine the critical parameters for the fingering instability, we use the interfacial length corresponding to $\Rc = 0$ as the reference. A set of parameters ($Da, \Rc$) is assigned an unstable displacement if the corresponding interfacial length exceeds more than $5\%$ from the reference value; otherwise it is assigned a stable displacement. Thus obtained stable and unstable cases are summarized in figure 
\bfig
\centering
\includegraphics[scale=1]{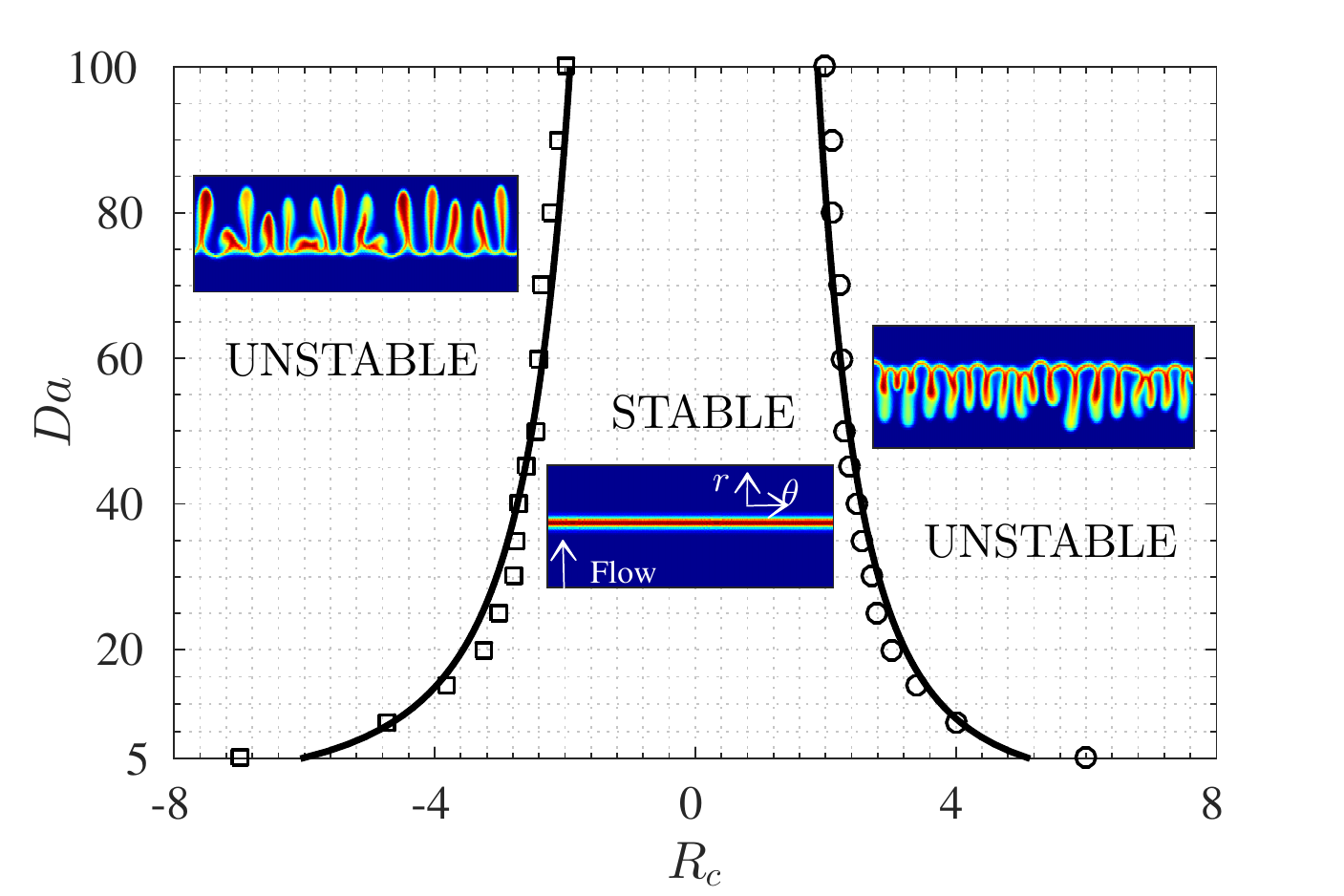} 
\caption{$Da-\Rc$ parameter space showing stable as well as unstable regions for $Pe = 3000$. $\protect \sqre$
denote $\Rm$ and $\protect \crcl$ denote $\Rp$, while the solid lines represent the power laws $\Rp= 8.8318 Da^{-0.3365}$ and $\Rm= -11.2329 Da^{-0.3833}$. The truncated density plots of product concentration in $(r, \theta)$ coordinates representing each zone are also shown in the inset.}
\label{fig:DA_RC_plane} 
\efig
\bfig
\centering
\includegraphics[scale=0.65]{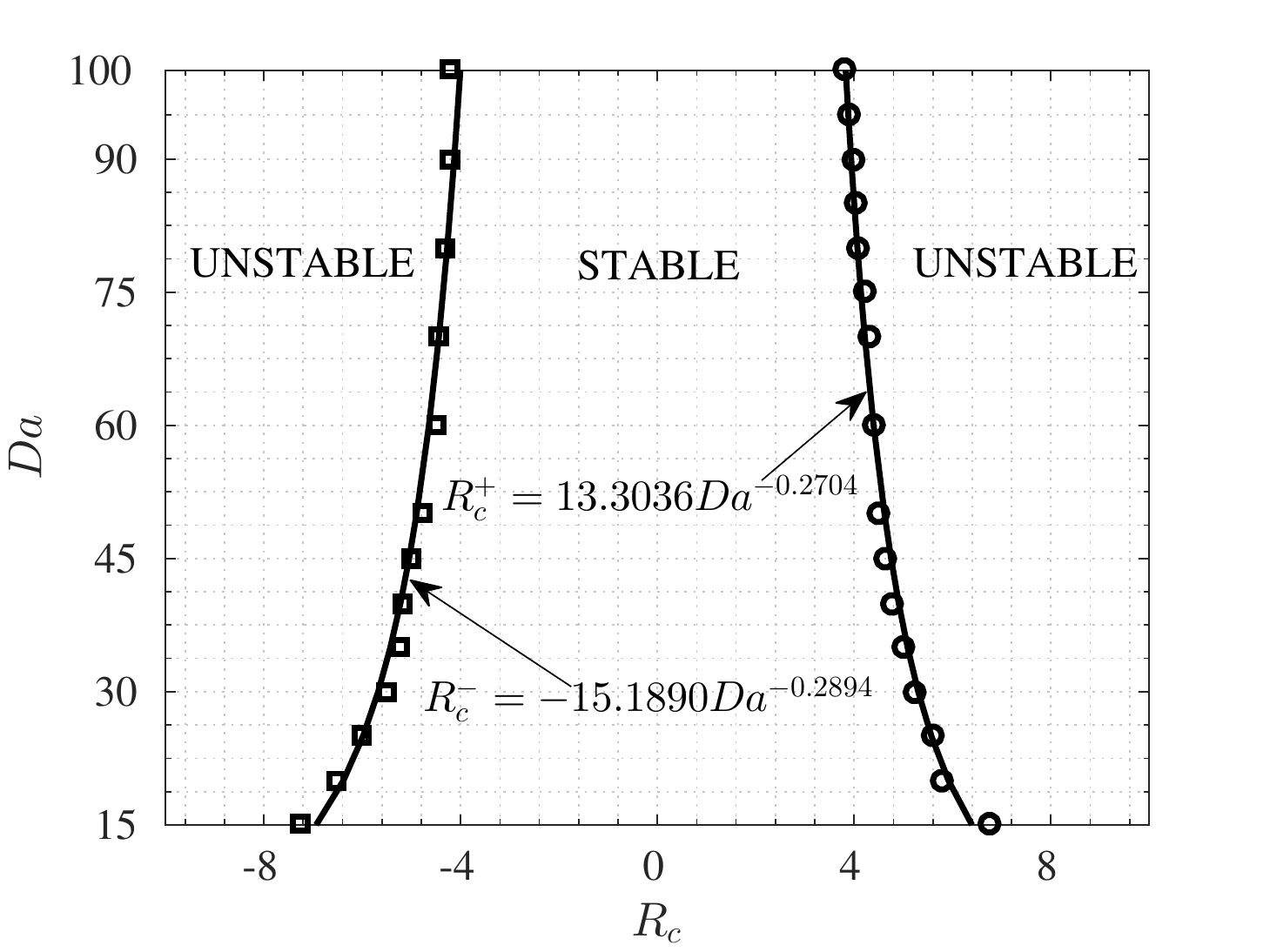} 
\caption{ Widening of the stable region on decreasing $\Pen$.
$Da-\Rc$ parameter space for $Pe = 1000$ with the solid lines representing the power laws $\modu{\Rc^{\pm}} =\alpha_{\pm}  Da^{-\beta_{\pm}}$. Critical $\Rc$ is shown by symbols with $\protect \sqre$ and $\protect \crcl$  respectively
denoting $\Rm$ and$\Rp$.}
\label{fig:DA_RC_plane_1000} 
\efig
\ref{fig:DA_RC_plane} for $\Pen = 3000$, which depict that the $Da-\Rc$ plane consists a stable region around $\Rc = 0$, sandwiched between two unstable regions. This stable region exists for all $Da$ values considered and narrows as $Da$ increases. Therefore for every finite $Da$, we captured critical values of $\Rc$ denoted as $\Rm$ and $\Rp$, such that no VF occurs when $\Rm \leq \Rc \leq \Rp$ and VF is observed outside this interval. Symbols in figure  \ref{fig:DA_RC_plane} denote the critical parameters $\Rm$ (squares)$,\Rp$ (circles), while the lines represent a least squares fit to the function $\modu{\Rc^{\pm}} \propto Da^{-\beta_{\pm}}$. A naive observation to be made here is that a higher viscosity gradient is required to trigger the instability for the reaction producing a less viscous product as $|\Rm| > |\Rp|$ for all $Da$. The numerical simulations are carried out for a wide range of $\Pen$ and the qualitative results presented in figure \ref{fig:DA_RC_plane} remain unaffected for other values of $\Pen$ as well, as evident from the $Da-\Rc$ parameter space for $\Pen =1000$ in figure \ref{fig:DA_RC_plane_1000}. It is verified that the stable region reduces as $\Pen$ increases, but never completely disappears.

Now, we give a relation between the critical parameters so as to mathematically determine the bounds of the stable region for any given moderate $Da$. \citet{Bischofberger2014} experimentally determined the existence of critical viscosity contrast ($R_{crit}$) for the fingering instability to occur in the miscible VF. Following the stability analysis of fingering instability, we can write the critical viscosity ratio as $\mcrit \approx e^{R_{crit}}$. We extend this idea for the current case of reactive fingering instability in miscible fluids. We write the critical viscosity $\mcrit^{\pm}$ as $e^{\modu{\Rc^{\pm}} f(Da)}$, where $f(Da)$ is a function of $Da$ to be determined. Setting $\mcrit^{\pm} = \mcrit$ and using a best fit of the numerically obtained critical parameter, we obtain \beq
\label{eq:powerlaw_fit}
\Rp = \alpha_{+} Da_c^{-\beta_{+}} , \qquad \Rm  = -\alpha_{-} Da_c^{-\beta_{-}}, 
\eeq
where $\alpha_{\pm}$ and $\beta_{\pm}$ are the fitting parameters which depend on $\Pen$. These parameters for $\Pen=3000,1000$ are given in figure \ref{fig:DA_RC_plane} and \ref{fig:DA_RC_plane_1000}, respectively. Clearly the stable region widens on decreasing $\Pen$. This clearly is a consequence of the stabilising diffusive force on miscible VF. Thus, for a given $\Pen$, we can divide the $Da-\Rc$ parameter space into three regions which can be used for suitably choosing the parameters to control VF.

\section{Conclusion}\label{sec:conclusion}
With an aim to explore the effect of the chemical reaction and geometry on VF, we numerically studied the dynamics of chemical reaction-induced fingering instabilities in miscible radial displacements. We choose a pair of viscosity matched reactants, such that in the absence of chemical reaction one stably displaces other. Thus, chemical reaction induces a non-monotonic viscosity variation radially.  We have explored a range of parameter values covering a wide spectrum of reaction rate and the cases of viscosity increasing as well as viscosity decreasing chemical reactions. Depending on the viscosity of the end product, either the inner or the outer interface of the product ring becomes unstable, provided the viscosity contrasts between the displacing and the displaced fluid is larger than a critical value. The instability of the inner and outer interfaces of the product is summarized in a parameter space spanned by $\Rc$ and $Da$. Our results show that a non-zero viscosity contrast is not sufficient for VF to occur, which is also true for classical miscible fingering instability in non-reactive fluids \citep{Bischofberger2014}. We have shown that fingering instability occur when the mobility ratio reaches beyond a critical value, which depends on both $Da$ and $\Pen$. Depending on these critical parameter values, for a fixed $\Pen$, the $\Rc$-$Da$ parameter space is divided into three regions--one stable region separating two unstable regions. For a given $\Pen$, the critical log mobility ratios and the critical  Damk\"ohler number are found to follow a power law. The stable region exists for all $\Pen$, with its the width being dependent on $\Pen$. 

Fingering dynamics observed in the unstable regions of the $Da-\Rc$ parameter space are in good qualitative agreement with the experiments. Besides triggering VF, the chemical reaction also introduces a barrier to the finger growth. A delayed onset of VF for $\Rc < 0$ in comparison to that for $\Rc > 0$ is a resultant of the coupled effect of the stable barrier and a decaying $\upot$. This is exactly opposite to the predictions from the linear stability in a rectilinear flow \citep{Hejazi2010a}. Thus, there is an essence to consider more theoretical and numerical studies on miscible reactive VF in a radial geometry. Furthermore, our numerically predicted $Da-\Rc$ parameter plane can be accessed to controlling VF by suitably choosing the reactants and the reaction. We believe more experiments will be conducted to improve the numerically predicted boundary between the adjacent regions. 

\section*{Acknowledgements}
MM acknowledges the financial support from SERB, Government of India through project grant number MTR/2017/000283. 
SP acknowledges the support of the Swedish Research Council Grant No. 638-2013-9243. C.Y.C  is thankful to ROC (Taiwan) Ministry of Science and Technology, for financial support through Grant No. MOST 105-2221-E-009 -074-MY3. VS and MM thank  Y. Nagatsu for  various fruitful discussions. VS also acknowledges 2017 NCTU Taiwan Elite Internship Program for the financial support to visit C.Y.C.



\end{document}

%% file: macros
\def \Rp {R_c^+}

\def \bx {\boldsymbol{x}}

\def \Rm {R_c^-}
\def \Rc {R_c}

\def \mcrit {\mu_{\rm crit}}
 
\def \DC {D_C}
\def \DA {D_A}
\def \DB {D_B}

\def \mc {\mu_C}

\def \ms {\mu^{\ast}}
\def \bu {\boldsymbol{u}}
\def \upot {\boldsymbol{u}_{\rm pot}}
\def \urot {\boldsymbol{u}_{\rm rot}}
\def \pa {\partial}
\def \om {\omega}

\def \ro {r_{\rm out}} 
\def \roo {r_{0, {\rm out}}}
\def \s {\sigma}
\def \a {\widetilde{a}}
\def \b {\widetilde{b}}
\def \c {\widetilde{c}}
\def \u {\widetilde{\bu}}
\def \p {\widetilde{p}}
\def \m {\widetilde{\mu}}
\def \rin {r_{\rm in}} 

\def \r {r_{0, {\rm in}}} 
\def \wt {\widetilde}

\def \tt {\wt{t}}

\def \s0 {\s_0}

\def \bseq {\begin{subequations}}
\def \eseq {\end{subequations}}
\def \bseqn {\begin{subeqnarray}}
\def \eseqn {\end{subeqnarray}}

\def \lap {\nabla^2} 
\def \beq {\begin{equation}}
\def \eeq {\end{equation}}
\def \bfig {\begin{figure}}
\def \efig {\end{figure}}
\def \beqn {\begin{eqnarray}}
\def \eeqn {\end{eqnarray}}
\def \k {\kappa}
\def \Lp {L^+}
\def \Lm {L^-}
\def \nn {\nonumber}
\def \a {\alpha}

\newcommand{\abra}[1]{\left\langle #1\right\rangle}

\newcommand{\bra}[1]{\left( #1 \right)}
\newcommand{\sqbra}[1]{\left[ #1 \right]}

\newcommand{\grad}[1]{\bnabla #1}
\newcommand{\tgrad}[1]{\widetilde{\bnabla} #1}
\newcommand{\tdiv}[1]{\widetilde{\bnabla} \bcdot #1}
\newcommand{\divg}[1]{\bnabla \bcdot #1}

\newcommand{\modu}[1]{\left\lvert #1 \right\rvert}
\newcommand{\crcl}{\raisebox{0.6pt}{\tikz{\node[draw,scale=0.5,circle,fill=none](){};}}}
\newcommand{\sqre}{\raisebox{0.5pt}{\tikz{\node[draw,scale=0.5,regular polygon, regular polygon sides=4,fill=none](){};}}}